\documentclass[twocolumn,floatfix]{aastex63}
\usepackage{hyperref}
\usepackage{xspace} 
\xspaceaddexceptions{]\}}

\newcommand{\batse}{{\em BATSE}\xspace}
\newcommand{\fermi}{{\em Fermi}\xspace}
\newcommand{\swift}{{\em Swift}\xspace}

\begin{document}

\title{Classification of BATSE, Swift, and Fermi Gamma-Ray Bursts from Prompt Emission Alone}

\correspondingauthor{Charles Steinhardt}
\email{steinhardt@nbi.ku.dk}

\author[0000-0003-3780-6801]{Charles L. Steinhardt}
\affiliation{Cosmic Dawn Center (DAWN)}
\affiliation{Niels Bohr Institute, University of Copenhagen, Lyngbyvej 2, DK-2100 Copenhagen \O}

\author[0000-0001-7691-0762]{William J. Mann}
\affiliation{Cosmic Dawn Center (DAWN)}
\affiliation{Niels Bohr Institute, University of Copenhagen, Lyngbyvej 2, DK-2100 Copenhagen \O}

\author[0000-0001-7633-3985]{Vadim Rusakov}
\affiliation{Cosmic Dawn Center (DAWN)}
\affiliation{Niels Bohr Institute, University of Copenhagen, Lyngbyvej 2, DK-2100 Copenhagen \O}

\author[0000-0002-8896-6496]{Christian K. Jespersen}
\affiliation{Department of Astrophysical Sciences, Princeton University, Princeton, NJ 08544, USA}

\begin{abstract}
Although it is generally assumed that there are two dominant classes of gamma-ray bursts (GRB) with different typical durations, it has been difficult to unambiguously classify GRBs as short or long from summary properties such as duration, spectral hardness, and spectral lag.  Recent work used t-distributed stochastic neighborhood embedding (t-SNE), a machine learning algorithm for dimensionality reduction, to classify all \swift~gamma-ray bursts as short or long.  Here, the method is expanded, using two algorithms, t-SNE and UMAP, to produce embeddings that are used to provide a classification for the 1911 \batse~bursts, 1321 \swift~bursts, and 2294 \fermi~bursts for which both spectra and metadata are available.  Although the embeddings appear to produce a clear separation of each catalog into short and long bursts, a resampling-based approach is used to show that a small fraction of bursts cannot be robustly classified.  Further, 3 of the 304 bursts observed by both \swift~and \fermi~have robust but conflicting classifications.  A likely interpretation is that in addition to the two predominant classes of GRBs, there are additional, uncommon types of bursts which may require multi-wavelength observations in order to separate from more typical short and long GRBs.
\end{abstract}

\section{Introduction}
\label{sec:intro}
The most prominent feature of a $\gamma$-ray burst (GRB) is the short duration of its prompt emission, ranging from $\sim 10^{-2}$s to $\sim 10^3$s.  The distribution of observed durations is predominantly bimodal, leading to a standard division of GRBs into {\em short} and {\em long} bursts (cf. \citet{kouveliotou1993}).  These two groups have been hypothesized to have distinct astrophysical origins, with short bursts associated with mergers of neutron stars \citep{Tanvir2013,berger2013,Ghirlanda2018} and long bursts associated with core collapses of massive stars \citep{hjorth2003,Stanek2003}.  This would imply that it should be possible to cleanly separate the two types of bursts based on observed properties.

However, there is considerable overlap between the two distributions in duration.  Thus, the standard dividing line at $T_{90} = 2$s will miscategorize the shortest long bursts and the longest short bursts \citep{tavani1998,paciesas1999}.  The most prominent additional observed features, spectral hardness \citep{kouveliotou1993} or spectral lag \citep{Norris1986,Norris2006} do not provide a clean separation between short and long bursts. Environment has also been considered as a possible distinguishing factor \citep{Lesniewska22}. More complex derived properties, or combinations of the standard summary properties, have also been unable to provide a complete separation \citep{06environmentfruchter,NAKARsgrb,Bromberg11,zhang2012revisiting,Bromberg_2013classification}.

Recently, a new approach using the dimensionality reduction algorithm t-distributed Stochastic Neighbor Embedding (t-SNE) used the full \swift light curves to provide the first clear separation between short and long bursts \citep{Jespersen2020}. 
 Although the same approach should be expected to successfully separate bursts from the \batse and \fermi catalogs as well, the differences between these datasets require individualized tuning and the datasets cannot be combined into one t-SNE map without significant information loss.  Thus, although this work uses very similar methodology to the \citet{Jespersen2020} pilot study, the methodology applied to the other two catalogs is necessarily slightly distinct.

This work expands upon that study in several significant ways: \begin{itemize}
    \item{Bursts from \batse~and \fermi~are also mapped, providing a complete t-SNE-based classification catalog combining all large datasets.}
    \item{It is shown that the three catalogs yield similar separations, implying that the separation is truly astrophysical rather than due to selection effects, data reduction choices, or other systematics.}
    \item{An additional algorithm, Uniform Manifold Approximation and Projection (UMAP; \citealt{McInnes2018}) is considered as an alternative to t-SNE.  UMAP and t-SNE are both dimensionality algorithms, and often produce similar results when well-tuned \citep{Kobak2019,Becht2019,Xiang2021}.  However, for most large, high-dimensionality datasets, UMAP produces these results in significantly shorter computation time \citep{Hu2019}\footnote{See also \url{https://umap-learn.readthedocs.io/en/latest/performance.html}}.}
    \item{A comparision of the classifications of bursts which appear in both the \swift~and \fermi~datasets is used both as a consistency check and to determine whether additional subclassifications are suggested by the t-SNE and UMAP maps.}
    \item{A new measure of uncertainty is introduced to describe the stability of these classifications.}
\end{itemize}

In \S~\ref{sec:method}, the methods used in \citet{Jespersen2020} are reviewed, modified as needed, and applied to the \batse~and \fermi~catalogs.  Corresponding choices for UMAP are also discussed.  The resulting classifications are presented in \S~\ref{sec:catalogs}.  In \S~\ref{sec:cross-match}, the robustness of this classification is evaluated using a combination of multiple catalogs.  The results and implications for future surveys are discussed in \S~\ref{sec:discuss}.  A full catalog including classifications for objects in \batse~, \fermi~, and \swift~is included, as described in the Appendix.

\section{Methodology}
\label{sec:method}

The methodology used in this work follows the general approach used in \citet{Jespersen2020} analysis of the \swift dataset, modified in order to accommodate the differences between various GRB observatories.  Dimensionality reduction is applied to the full set of light curves in every observed band.  The resulting embedding is examined, and divided into cleanly-separated structures.  The objects in each structure are then considered to comprise a distinct class of GRBs.

In practice, there are several additional steps.  First, the data must be standardized, removing incomplete or missing observations.  Then, preprocessing is performed to remove or negate irrelevant information that these algorithms might otherwise interpret as meaningful.  Afterwards, dimensionality algorithms can be applied, and that application requires the choice of several hyperparameters which must be selected individually for each dataset.  Finally, it is necessary to determine which structures on the resulting embedding should be considered distinct.  Within the same approach, each of these steps requires different choices for \swift, \batse, and \fermi.  These are described in more detail in the subsections below.

It is important to note that ``unsupervised'' algorithms such as the ones used here are in practice strongly dependent on the choice of hyperparameters.  The proper interpretation of the embeddings presented here should not be that the hyperparameters we have chosen are correct and others incorrect.  Rather, every choice of hyperparameters leads to a valid embedding, which contains potentially useful information about the distribution but is always incomplete.  In that sense, it would be more like considering various projections.  Unlike projections, however, there is no rigorous formalism such as principal component analysis for determining which will be most useful.

Here, the choices rely on the physical assumption that GRBs can be divided into discrete groups due to distinct progenitors, but avoid asserting any specific number of progenitors. Rather, the hyperparameters which produce the cleanest separations are selected.  In that respect, the separation into two primary groups is a property of the dataset.  However, because these embeddings focus on groups of a specific size, additional progenitors which are less common and thus have significantly fewer examples will not be revealed on these maps.  

Emebeddings focusing on small groups were considered in \citet{Jespersen2020}, but did not produce easily-interpretable, well-separated groups.  Further, as described in \S~\ref{subsec:crossmatch}, the smaller substructures hinted at on the embeddings considered here do not appear to be meaningful.  However, given the large space of possible hyperparameters, of which only a small portion has been sampled in this work, it is likely that additional, astrophysically-interpretable groups could exist.  Identifying how many and which of these groups are meaningful might require additional observaional information.

\subsection{Burst Selection}

One of the restrictions of t-SNE and UMAP is that they cannot be applied to data of different dimension or labels.\footnote{Or, at least, the distance metric selected must make a choice about how to handle the difference between missing and measured dimensions in a consistent way, which is typically very difficult unless the correct answer is already known.}  This is one of the reasons that \swift, \batse, and \fermi are examined individually rather than as a group; the different bands and cadences of each set of observations do not allow a direct comparison.  Here, bursts which at least one band is predominantly missing or key metadata do not exist are rejected entirely.  More minor flaws are generally adjusted or corrected instead of rejecting the burst.  The specific choices for each individual dataset are included for reproducibility.

For \batse \citep{BATSEcatalog}, the ASCII file is used as the canonical source of each light curve, due to the similar \texttt{tte\_bfits} files having various data problems.  Out of 2702 total bursts, 527 bursts with missing ASCII files on the HEASARC server{\footnote{\url{https://heasarc.gsfc.nasa.gov/FTP/compton/data/batse/ascii_data/64ms/}}} were therefore rejected.
Summary data including flux, fluence and T90 were taken from \url{https://batse.msfc.nasa.gov/batse/grb/catalog/current/}.  An additional 264 bursts with missing summary data were rejected.  In total, 791 bursts were rejected and the remaining 1911 were included.

For \swift \citep{Lien2016}, the same choices were made as in \citet{Jespersen2020}. This work uses a more recent version of the catalog, which includes an additional 67 bursts.

For \fermi \citep{VonKienlin2020}, the \texttt{bcat} file is used as the canonical source of each light curve.  All but 34 bursts available at \url{https://heasarc.gsfc.nasa.gov/FTP/fermi/data/gbm/bursts/} had a \texttt{bcat} file available, so 3108 were downloaded.  The \fermi GBM Burst Catalog includes multiple models, and the best-fit average flux was used for each burst. These fluxes and t90 were taken from \url{https://heasarc.gsfc.nasa.gov/W3Browse/fermi/fermigbrst.html}. Unfortunately, nearly all of the bursts recorded after mid-2018 had not undergone spectral analysis, therefore lacking best-fit flux and fluence models.  These 814 bursts were cut, leaving 2294 remaining for analysis. 

A potential concern is that the use of deconvolved flux lightcurves for \fermi but count rate lightcurves for the other observatories makes the three no longer directly comparable.  However, this was already the case because, e.g., \swift and \fermi provide different bands and thus different information about each bursts.  Using different data types can be thought of as simply another difference between datasets and pipelines.  As shown in \ref{sec:cross-match}, there is still strong agreement between the \swift and \fermi classifications, indicating that all of these differences between observatories, reductions, and choice of lightcurves do not significantly alter the conclusions presented here about the broad classification of GRB.

\subsection{Preprocessing}

Several preprocessing steps are required before dimensionality reduction algorithms can be used.  Because these algorithms require identical formats and have difficulty handling missing information, light curves are padded with additional zeros to produce data of identical lengths as in \citet{Jespersen2020}.  The goal of additional preprocessing is to remove extraneous information which might otherwise contribute to the post-embedding positions of light curves while retaining as much information as possible.  There are two key extraneous parameters: the overall brightness of the burst and the trigger time.

Although the energy carried by a burst is physically meaningful, for most GRBs in the catalog, the redshift is unknown.  Under the assumption that bursts with the same underlying astrophysical origins can exist at a wide range of redshift, the brightness will be a poor indicator of luminosity.  Therefore, each burst is normalized by dividing by the total fluence.\footnote{For \fermi, we instead normalized by average flux, which we found to work better on the semi-processed \texttt{bcat} data.}  This retains hardness information because the relative brightness between bands is preserved.

The other issue is handling trigger time offsets, where the time that the burst was detected differs from the actual start of the burst.  These offsets are typically due to instrumentation rather than due to the shape of the burst itself, and therefore should be discarded.  To accomplish this, the same procedure is used as in \citet{Jespersen2020}.  A discrete-time Fourier transformation (DTFT) is performed on a concatenation of the lightcurves in all observed bands.  An overall time shift will only change the phase information under this DTFT.  However, relative time offsets between different bands, as well as other meaningful information such as duration, hardness, and spectral lag, will all contribute to the amplitudes as well.  Therefore, the phase information is then discarded, and dimensionality reduction algorithms are run only on the amplitudes.

\subsection{Embedding}

Dimensionality reduction algorithms are then run on each of the datasets independently.  It is necessary to perform different embeddings for \swift, \batse, and \fermi, since they measure different bands with different cadences and report data in different formats.  Two main algorithms are used here: t-SNE and UMAP.  Both have hyperparameters which must be tuned for each individual dataset.  The primary hyperparameter of importance for t-SNE is perplexity.  For UMAP, there are two hyperparameters which must be tuned, n\_neighbors and set\_op\_mix\_ratio.

\begin{table}[!ht]
\begin{center}
 \begin{tabular}{lccc} 
 \hline
 Dataset & t-SNE & UMAP & UMAP \\  
  & perplexity & n\_neighbors & set\_op\_mix\_ratio \\ 
 \hline
 \batse & 0 & 0 & 0.25  \\
 \fermi & 20 & 25 & 0.25  \\
 \swift & 20 & 20 & 0.3  \\
 \hline
\end{tabular}
\caption{Hyperparameters chosen for the t-SNE and UMAP embeddings of each of the three GRB catalogs used in this work. }
\label{tab:hyperparameters}
\end{center}
\end{table}

In every case, a range of embeddings with different properties will result from various choices of hyperparameters.  Unfortunately, there is no rigorous mathematical formalism for choosing optimal hyperparameters.  The multiple embeddings which can result from different choices are in some sense all correct and valid, yet simultaneously are all incomplete descriptions of the full structure.  Naturally, some will be more useful for GRB classification.  For each dataset, both t-SNE and UMAP were run with a variety of hyperparameters, and the ones which produced embeddings with the cleanest separations were chosen.  The hyperparameters chosen are summarized in Table \ref{tab:hyperparameters}.  

Although these embeddings often show clear separations between short and long GRB, that does not guarantee that there are only those two classes of GRB.  The \texttt{perplexity} and \texttt{n\_neighbors} hyperparameters for t-SNE and UMAP, respectively, essentially dictate the size of the groups which the embeddings focus on representing properly.  Thus, if there are two predominant groups and one or more tiny ones, the tiny groups might be attached to a larger one.  An attempt to identify bursts which might not be standard short or longer GRBs is described in \S~\ref{subsec:resampling}.

\section{Classification}
\label{sec:catalogs}

For each catalog, two embeddings are produced, one with t-SNE and the other with UMAP, for a total of six maps.  On each map, a division is made into two large groups, labeled short and long based on the typical duration in each group.  A small fraction of bursts are classified either as outliers, distinct from both groups, or as ambiguous, lying in the region between the short and long groups.  The \batse~map is more complicated, producing an initial separation which produces a duration distribution qualitatively different than the other two catalogs.  A more quantitative comparison is performed in \S~\ref{sec:cross-match}.

\subsection{Swift}

\begin{figure}[!ht]
    \centering
    \includegraphics[width=0.5\textwidth]{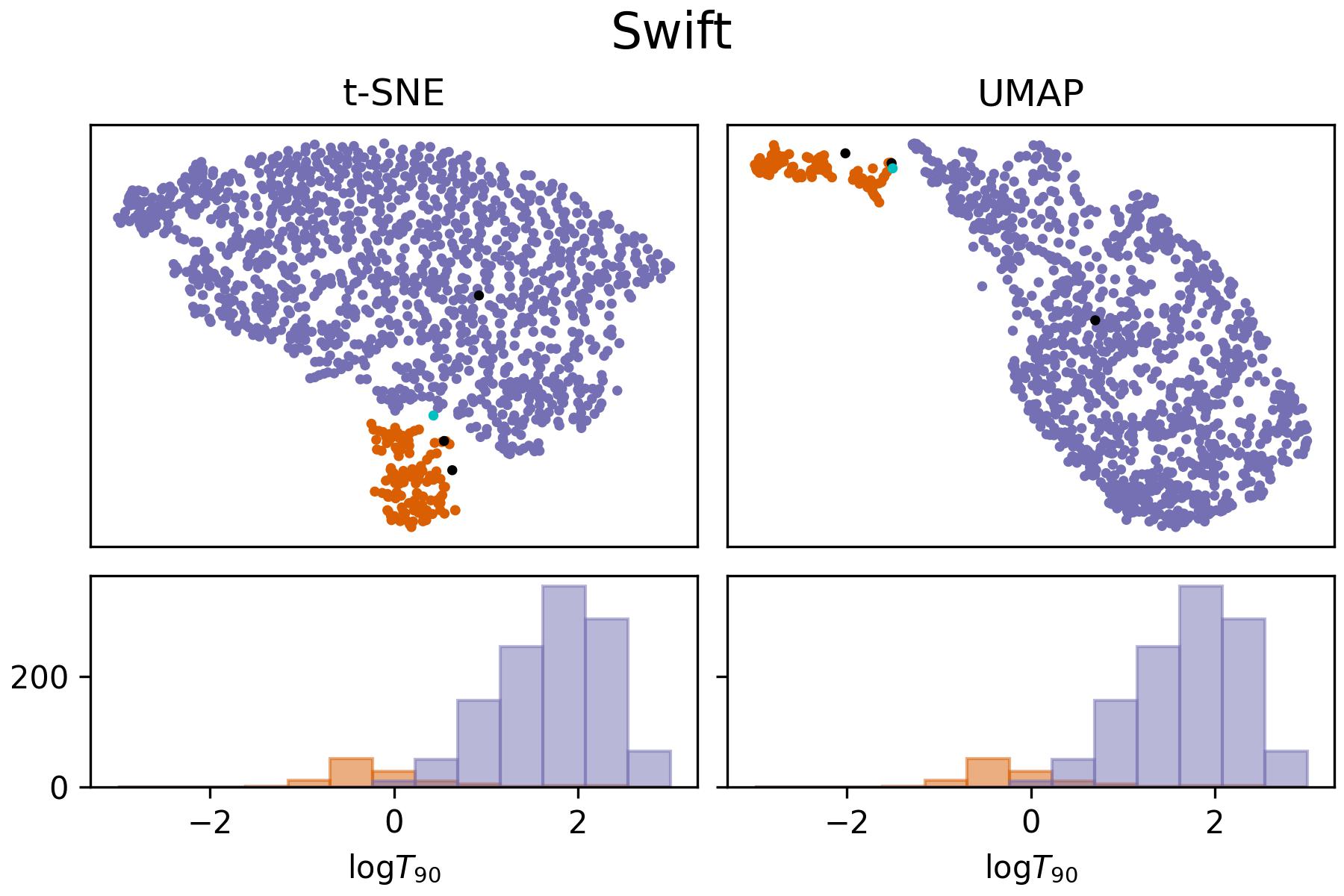}
    \caption{t-SNE (left) and UMAP (right) embeddings of 1321 \swift~lightcurves, colored by classification.  The duration distributions (bottom) are consistent with an interpretation as separation into short (orange) and long (purple) GRB rather than merely a classification purely by duration.  The two embeddings agree on the classification of all but one burst (cyan).  Three bursts are classified different in this work than in \citet{Jespersen2020} (black).}
    \label{fig:swift_map}
\end{figure}

Since the technique used in this work is very similar to that in \citet{Jespersen2020}, the embeddings and classifications produced using the \swift~catalog are nearly identical.  There is a clear separation into two groups. one identified as short (orange, Fig. \ref{fig:swift_map}) and the other as long (purple).

Of the 1321 bursts, t-SNE classifies 114 as short and 1207 as long.  The UMAP embedding classifies just one burst differently: GRB090813 (cyan, Fig. \ref{fig:swift_map}) is long in the t-SNE embedding but short in UMAP.  In addition, three bursts as classified differently here than in \citet{Jespersen2020}.  GRB121226A and GRB180418A are long in the Jespersen catalog and short here, while GRB050724 switches classification from short to long.

Although it may seem counterintuitive that a burst can switch classification when an identical technique is run on a superset of the data, this is indeed a property of both t-SNE and UMAP.  Both algorithms assign a cost to placing every pair of objects at any specific distance, such that similar objects are less costly at short distances and dissimilar objects less costly at large distances, then seek to minimize the global sum of that cost.  The addition of a new point will typically result in a lowest-cost configuration that involves not merely placing that point on an existing map, but shifting their locations as well.  For example, an analogous physical system might be one in which every object is attached to every other by a spring, with the stiffness of that spring depending upon their similarity.  The addition of a new set of springs will likely result in all of the distances changing in the equilbrium configuration.

One consequence of the choice of perplexity (for t-SNE) and n\_neighbors (UMAP) is that both embeddings attempt to place outliers into a cluster where plausible.  Thus, bursts which are somewhat dissimilar to both short and long GRBs are often placed on the edges of whichever group is more similar.  The addition of a small number of similar objects can therefore change which group they are located close to.  Thus, GRB050724, GRB121226A and GRB180418A are likely neither typical short bursts nor typical long bursts, but instead outliers, either for astrophysical reasons or due to a data processing artifact.  An attempt to identify similar bursts in other datasets is described in \S~\ref{subsec:resampling}.

\subsection{Fermi}

\begin{figure}
    \centering
    \includegraphics[width=0.5\textwidth]{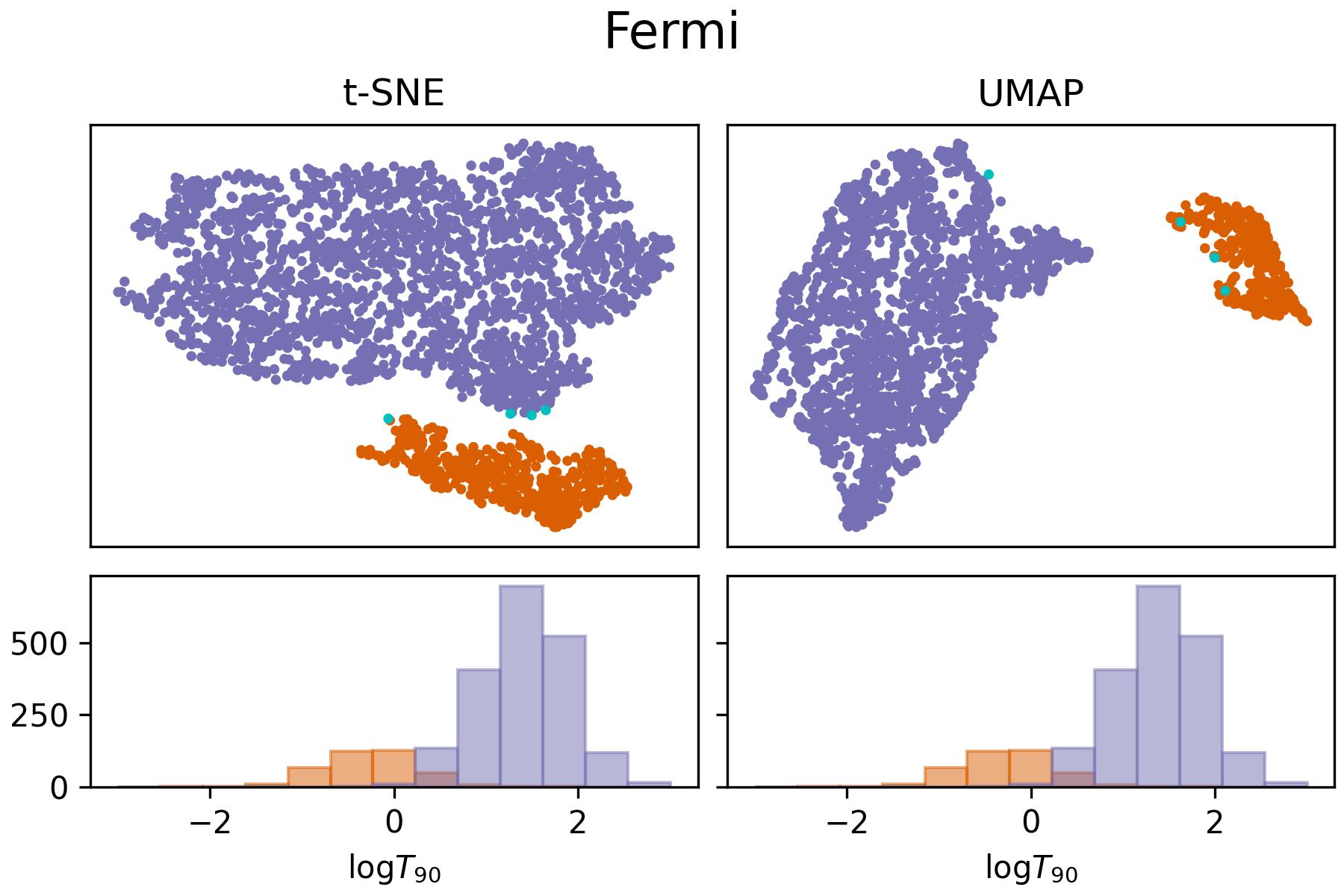}
    \caption{t-SNE (left) and UMAP (right) embeddings of 2294 \fermi~lightcurves, colored by classification.  The duration distributions (bottom) are consistent with an interpretation as separation into short (orange) and long (purple) GRB rather than merely a classification purely by duration.  Although both embeddings show a clear separation, four bursts (cyan) are classified differently by t-SNE and UMAP.}
    \label{fig:fermi_map}
\end{figure}

The 2294 bursts in the \fermi~catalog are arranged into the two embeddings shown in Fig. \ref{fig:fermi_map}.  These maps produce the clearest separation of any of the three datasets using both t-SNE and UMAP.  As a result, no bursts were unable to be clearly assigned to either group.  A possible interpretation is that the higher-energy bands in the \fermi~dataset are more useful for distinguishing between types of bursts than the bands available in \batse~and \swift.  Had \fermi observed the full set of \batse~and \swift bursts, under this interpretation the \fermi embedding would be expected to look nearly identical on this larger dataset.

The t-SNE map classifies 387 bursts as short and 1907 as long.  On the UMAP map, there are 385 short bursts and 1907 long bursts.  Three bursts (cyan, Fig \ref{fig:fermi_map}) are classified as short by t-SNE and long by UMAP: GRB090811696, GRB110719825, and GRB110728056.  GRB080828189 (also shown in cyan) is the sole burst classified as long by UMAP but short by t-SNE.  The remaining 2290 classifiable bursts agree in both analyses. 

Still, given the completeness of the separation on both the t-SNE and UMAP maps, it is perhaps surprising that four bursts change classification between the two embeddings.  Several possible causes of this reclassification are evaluated in \S~\ref{subsec:resampling}.  As a result of that analysis, in the catalog presented here, a measure of the uncertainty in classification is developed.  For the remainder of this section, bursts will be described as short or long based on their most probable classification and the apparently clear separations in the embeddings in Figures \ref{fig:fermi_map}-\ref{fig:batse_map}.  The catalog associated with these work includes not only the central values but also estimated likelihoods for these classifications.  In that catalog, 2.0\% of \fermi~bursts have between a 10\% and 90\% probability of being classified as short (or long).  Such objects are therefore labeled as ambiguous rather than as short or long.

\subsection{BATSE}

\begin{figure}
    \centering
    \includegraphics[width=0.5\textwidth]{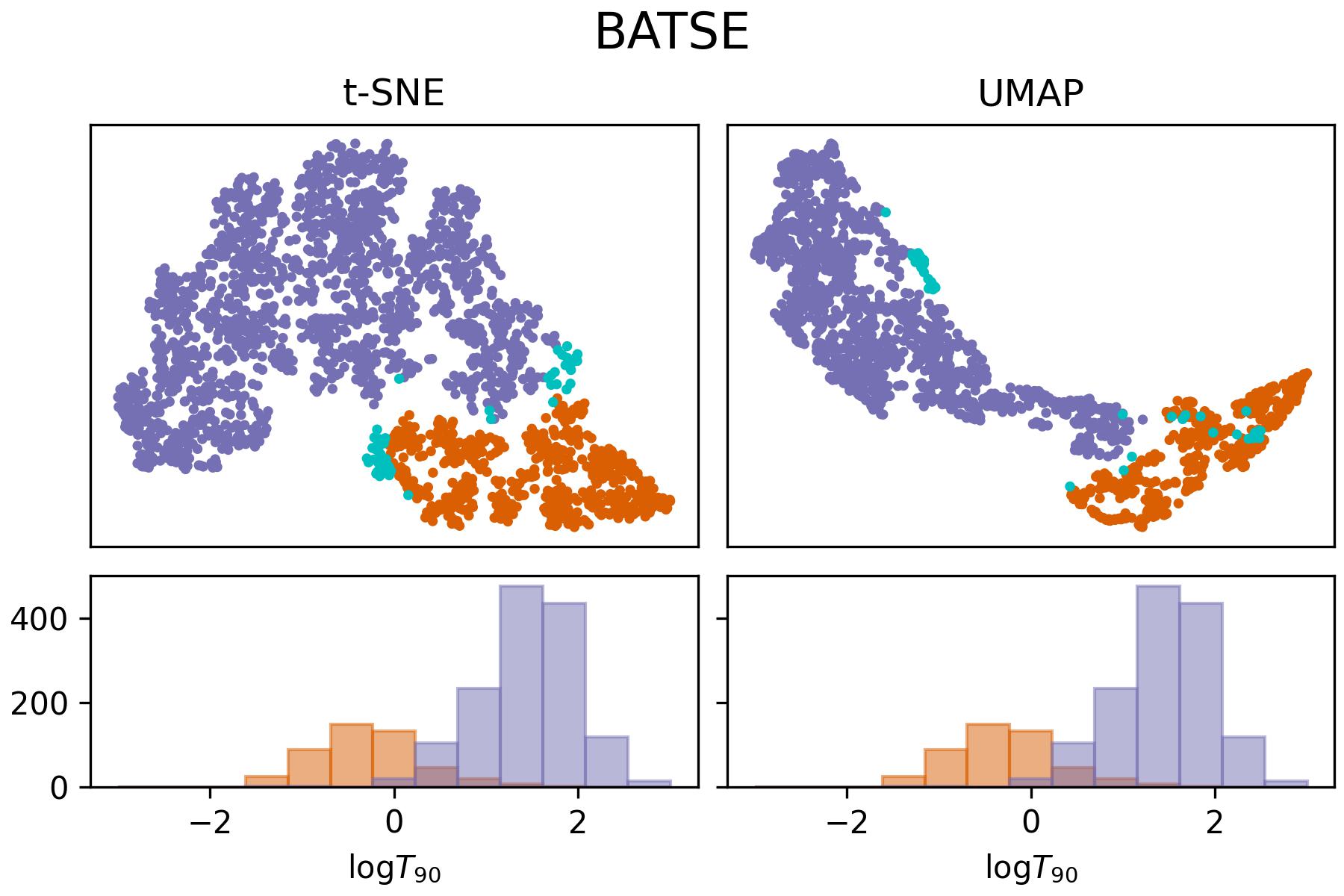}
    \caption{t-SNE (left) and UMAP (right) embeddings of 1911 \batse~lightcurves, colored by classification.  The duration distributions (bottom) are consistent with an interpretation as separation into short (orange) and long (purple) GRB rather than merely a classification purely by duration.  The separation in \batse~is less robust than the other datasets, likely due to higher noise, missing data or metadata, and the necessity to perform additional background subtraction.  As a result, 49 of the 1911 bursts (cyan) are classified differently by t-SNE and UMAP.}
    \label{fig:batse_map}
\end{figure}

The 1911 bursts in the reduced \batse~catalog are arranged into the embeddings shown in Fig. \ref{fig:batse_map}.  Unlike \fermi~and \swift, a significant number of \batse~bursts could not be included due to missing data, missing metadata, or high noise.  In total, 791 of the 2702 \batse~bursts were discarded, and many of the remaining ones have marginal quality and could not be well constrained.

The t-SNE embedding classifies 491 bursts as short and 1420 as long.  The UMAP embedding has similar size groups, with 484 long and 1427 long bursts.  However, unlike the \fermi~and \swift~embeddings, there is a more substantial disagreement in classification.  21 objects are classified as short by UMAP and long by t-SNE, and 28 classified as long by UMAP and short by t-SNE (cyan, Fig. \ref{fig:batse_map}).  The individual objects which switch classification between embeddings are indicated in the catalog associated with this paper.

The disagreement would have been far stronger using the \batse~light curves obtained directly from the HEASARC server.  To produce a separation, it was necessary to re-process each light curve in order to fit and subtract a background.  As a rudimentary subtraction, a linear fit was performed to the first (pre-burst) and last (well after $T_{100}$) data in each of the four bands, then subtracted from the full light curve.  The background-subtracted light curves were then used to produce the embeddings and catalog in this work.  

A more complete background subtraction would likely require rerunning or modifying the original processing pipeline.  It is likely that at the end of such an effort, an improved and more robust separation would be possible.  However, since higher-quality GRB data are available from newer observatories, here it is assumed that this would be of limited use.  Thus, in this work the decision was made to include this approximate background subtraction for completeness, but to focus on separating the \swift~and \fermi~datasets, as they will be most suitable for further analysis.

\section{Cross-matching and Validation}
\label{sec:cross-match}

A significant potential concern when using unsupervised machine learning methods is that because there is no training set, there is an inherent inability to validate the conclusions.  Although the GRB light curves can cleanly be separated into two groups, the methodology involved has no knowledge of astronomy or astrophysics.  Thus, the statement that there are two distinct classes of GRB light curves does not necessarily mean that there are two astrophysical mechanisms for producing GRB.  It could instead be that the two groups have been separated based on data artifacts or processing pipeline decisions.  t-SNE and UMAP are remarkable tools for finding clusters and categories, but not for determining the causes of those categories.   \citet{Jespersen2020} demonstrated that the \swift~classifications line up well with previous progenitor hypothesis.  For example, all \swift~bursts with a known, associated supernova afterglow were classified as long, consistent with previous expectations \citep{Hjorth2012,Cano2017}.  However, only a few \swift~GRB have observed afterglows, so it is difficult to rule out the possibility of this separation having been caused by data processing rather than astrophysical origin.

A key goal of this work is to combine observations from all three available GRB observatories, with different bands, sensitivity, selection, and processing pipelines.  If all produce a similar classification, this would validate the separation as being due to astrophysics.  Here, two tests are performed using multiple catalogs in an effort to determine whether this classification is robust.  

\subsection{Cross-matching Swift and Fermi GRB}
\label{subsec:crossmatch}

\begin{figure}[!ht]
    \centering
    \includegraphics[width=0.5\textwidth]{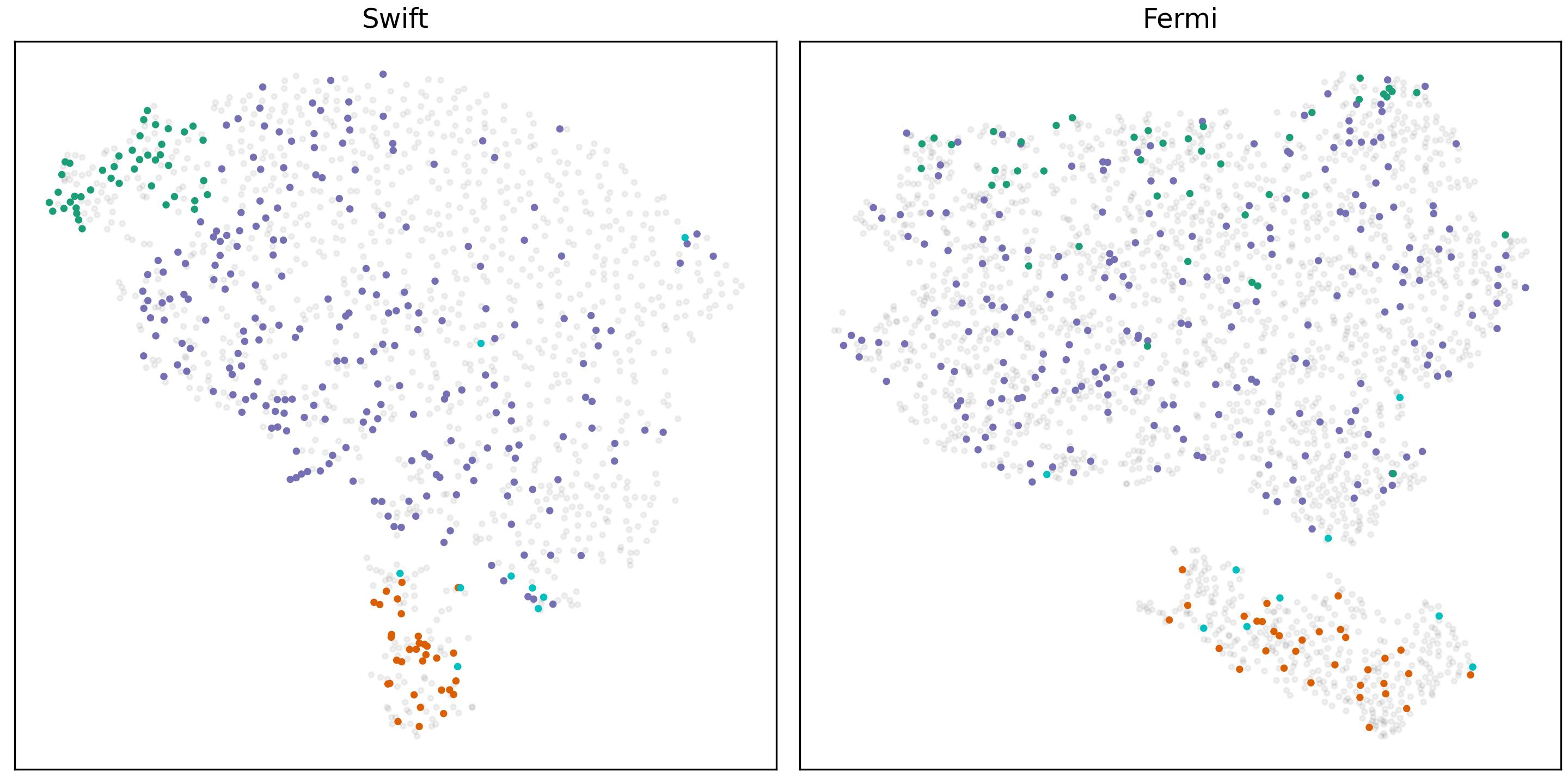}
    \caption{Locations of the 307 objects (various colors) common to both \swift~and \fermi~within the t-SNE embedding (gray).  298 of the 306 are classified in the same way for both datasets (orange for short; green or purple for long), which implies that the separation is due to the GRB emission rather than artifacts introduced in data reduction.  The remaining nine, which switch classification, are shown in cyan.  Although there are possible substructures on the t-SNE maps, such as in the upper-left corner of the \swift embedding in Fig. \ref{fig:swift_map}, these structures are not consistent between maps (green), and therefore are not interpreted as being of astrophysical origin.}
    \label{fig:comp_map}
\end{figure}

307 bursts were observed by both \swift~and \fermi, allowing a comparison between the two catalogs.  If the classification is robust and due to astrophysical origin, then bursts common to both catalogs must have the same label in both datasets.  Of the 307, 298 have the same classification and only 9 disagree (Fig. \ref{fig:comp_map}), which strongly suggests that the separation is indeed based on the emission itself rather than artifacts induced by the data reduction.  

The same approach also allows an investigation of the several smaller clusters which appear on t-SNE and UMAP maps.  The objects common to both telescopes which appear in a compact substructure of the long GRB groups (e.g., towards the top-left of the \swift~map in Fig. \ref{fig:swift_map}, shown as the green points on Fig. \ref{fig:comp_map}) do not comprise a distinct substructure in the other dataset, but rather are merely part of the long GRB group.   Therefore, these substructures are not interpreted as a distinct type of GRB or as having a distinct astrophysical origin, but rather as merely lying at one end of the parameter space of long GRBs.

\subsection{Classification Stability}
\label{subsec:resampling}

The 9 bursts classified differently in \swift~and \fermi~suggest that even the seemingly unambiguous separations in these embeddings might not be entirely robust.  Here, three sources of potential instability in classification are considered.  

\subsubsection{Data Ordering}

\begin{figure}
    \centering
    \includegraphics[width=0.5\textwidth]{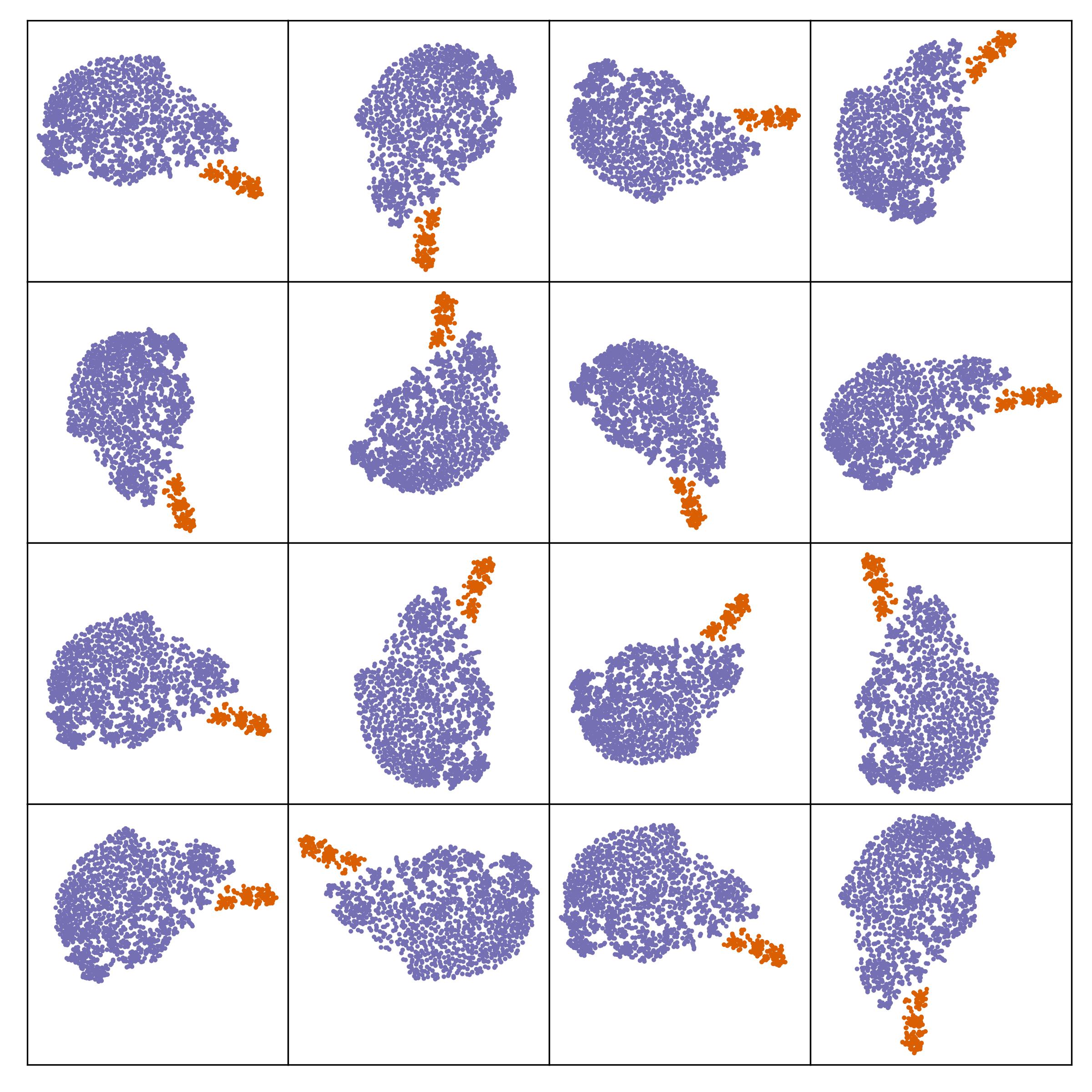}
    \caption{t-SNE embeddings of the \swift~GRB catalog using burst lists sorted randomly into 16 different orders.  On each map, short bursts are indicated in orange and long bursts in purple.  A comparison of 1000 maps generated for the \swift~dataset showed an identical classification for every object, confirming that the separation is robust to a change in the list order.}
    \label{fig:changeorder}
\end{figure}

First, the exact embeddings produced by t-SNE and UMAP depend upon the order in which the bursts are fed into the algorithm.  Although embeddings produced by different orders have almost identical structures, the actual locations of individual objects will vary (Fig. \ref{fig:changeorder}).  In order to investigate this, 1000 maps were generated for each of the three datasets using burst lists sorted randomly into different orders.  For each map, the bursts were divided into groups using spectral clustering \citep{Fiedler1973,Ng2001}, directed to split into exactly two groups.  In all 1000 trials, an identical set of short and long bursts was produced by both t-SNE and UMAP.  Thus, it can be concluded that the classification is resilient to changes in ordering.

\subsubsection{Outliers and Rare Bursts}

Perhaps a greater concern comes from the choice of hyperparameters and resulting handling of outlier bursts.  In this paper, the issue is described in terms of t-SNE hyperparameters, but is common to both algorithms.  Dimensionality reduction algorithms must choose between preserving more local and more global structure, as the data are not truly two-dimensional and some information must be lost in the mapping.  For t-SNE, this is controlled by the perplexity.  The embeddings here have been tuned to focus on separating the two major groups of GRBs.  

During the gradient descent as t-SNE iteratively optimizes its embedding, each burst is attracted by similar bursts and repelled by dissimilar ones.  For a perplexity of $N$, t-SNE imposes a probability density function which can be thought of as optimizing for the typical object having $N$ attractive neighbors.  This will produce a clean separation between two groups each larger than $N$.  However, tiny groups or individual objects with unique properties can be attached to the most similar group.  If a burst is, e.g., far more similar to short bursts than to long bursts, it will be classified as short.  However, if it has properties in common with both groups, then it will be attracted to both groups, and classified based on whichever set of attractors are stronger.  

In order to search for these groups, a resampling-based approach is adopted.  Subsets of (600, 900, 1000) bursts are drawn from the full (\swift, \batse, \fermi) catalog, corresponding to $\sim 50$\% of the total bursts, and an embedding produced for each subset.  As before, spectral clustering is applied to separate the embedding into two groups.  With the smaller samples, there are not always enough bursts to produce a clean separation without manual hyperparameter tuning.  Thus, only subsets which produce a clean separation similar to the original grouping are included.  Embeddings for which a Kolmogorov-Smirnov (KS) test indicates a $p < 0.10$ probability of being drawn from the same distributions as the separation on the full catalog are rejected\footnote{Note that by looking for similar distributions, this procedure estimates the stability of the specific bimodal classification being evaluated.  In principle, given computing resources well beyond those available to the authors, one could systematically search for the most stable bimodal classification.}.  A KS test is chosen rather than a test such as Anderson-Darling in order to emphasize the bulk of the distribution rather than the tails, so that embeddings which move outliers will not be excluded.

The hope is that if a burst is attracted by both groups, then it might change location.  In some of the random trials, many of its closest neighbors from one group or the other will be excluded.  However, a prototypical short burst will always be most similar to short bursts, even if some of its closest analogues are excluded.  

\begin{figure*}
    \centering
    \includegraphics[width=1\textwidth]{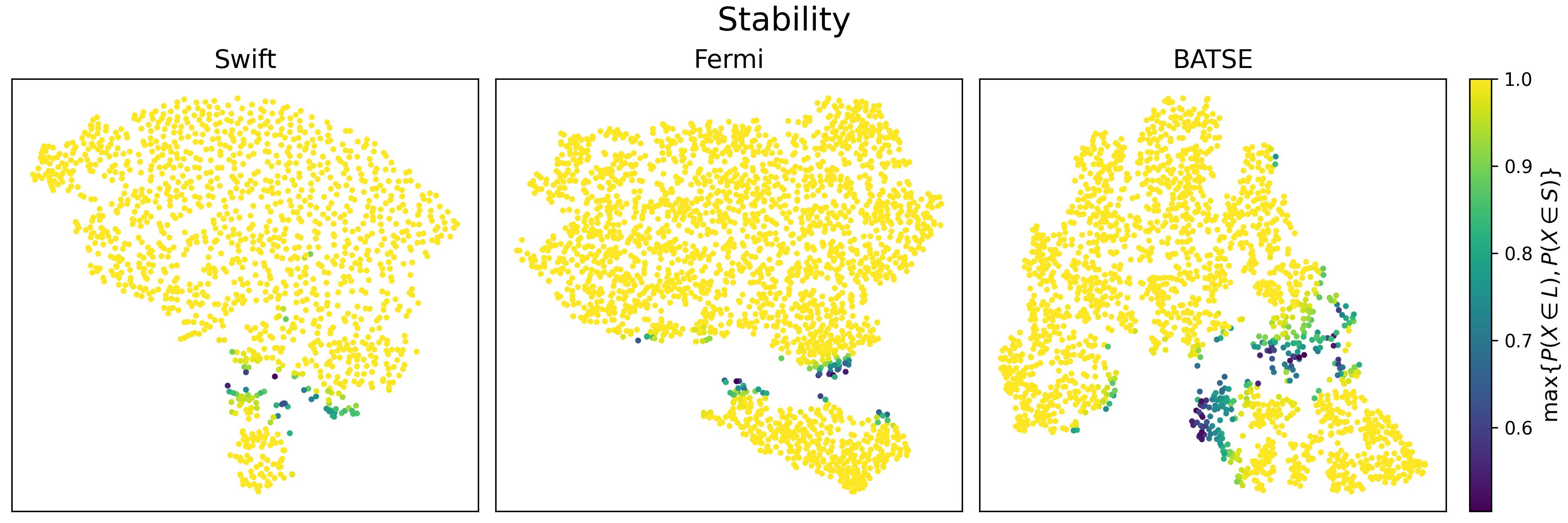}
    \caption {Stability of the classifications of bursts based on $\sim 750$ random subsets of 600, 1000, and 900 random bursts drawn from the \swift~(left), \fermi~(center), and \batse~(right) catalogs, respectively.  The vast majority of objects have an identical classification in every resampled embedding.  6.6\% of the \fermi~bursts change classification in at least one embedding, and 2.0\% are classified differently in at least 10\% of trials.  Objects in this last group are labeled as ambiguous rather than short of long in the catalog.  The fraction of ambiguous objects are larger for the \swift~(2.6\%) and \batse~(9.2\%) catalogs.}
    \label{fig:resampling}
\end{figure*}

For the most part, this procedure indicates that the classification is stable (Fig. \ref{fig:resampling}).  In the \fermi~catalog, the median resampling trial has 0.5\% of bursts change location.  6.6\% of bursts change location in at least one of the $\sim 750$ trials, and 2.0\% change location in more than 10\% of trials.  This latter group are labeled as ambiguous in the catalog from this work.  In the \swift~catalog, 0.5\% change location in the median trial, 13.0\% change location at least once and 2.6\% are ambiguous.  In the \batse~catalog, 2.8\% of bursts change location in the median trial, 20.6\% change location at least once and 9.2\% are ambiguous.

\subsubsection{Insufficient Information}

Of the 9 bursts classified differently by \swift~and \fermi, 7 change location in at least one resampled map in at least one catalog and 6 change location more than 10\% of the time (Table \ref{tab:outliers}).
\begin{table}[!ht]
\begin{center}
 \begin{tabular}{lcc} 
 \hline
 Burst & \swift & \fermi  \\  
 \hline
GRB090531B & 0.000 & 1.000  \\
 \hline 
GRB090927 & 0.070 & 1.000  \\
 \hline
GRB130716A & 0.000 & 1.000  \\
 \hline
GRB131004A & 0.194 & 0.754  \\
 \hline
GRB140209A & 0.682 & 0.081  \\
 \hline
GRB140320A & 1.000 & 0.346  \\
 \hline
GRB141205A & 0.257 & 1.000  \\
 \hline
GRB150120A & 0.146 & 0.860  \\
 \hline
GRB170318B & 0.926 & 0.000  \\
 \hline
\end{tabular}
\caption{Probability that one of the resampled maps will classify a burst as short for the 9 bursts with different classifications in the \swift~and \fermi~catalogs.  6 of the bursts have unstable classifications in at least one of the two catalogs, which implies that these are neither typical short nor long bursts.  One additional burst, GRB090927, changes location in 7\% of resampled \swift~maps.  However, two, GRB090531B and GRB130716A, have entirely stable but conflicting classifications.  These are likely extended emission bursts, and indicate that there is not enough information in either dataset alone to determine that they are atypical.}
\label{tab:outliers}
\end{center}
\end{table}
The remaining bursts, GRB090531B and GRB130716A, are consistently classified differently, in this case as long bursts by \swift~and short bursts by \fermi.  That is, in the \swift~dataset alone, they are similar to typical long burst and in the \fermi~dataset alone, they are similar to typical short bursts.  A reasonable interpretation is that these are extended emission bursts \citep{Norris2006,Kaneko2015}, which are known to be shorter in the harder emission observed by \fermi~and longer in \swift.  

The broader implication is that for some uncommon types of bursts, the information provided by one dataset alone is insufficient to classify them.  Extended emission bursts might only be detectable with a combination of harder and softer emission, which currently no single telescope can provide.  These bursts do appear distinct in an analysis which combines both \swift~and \fermi.  However, such a combination only exists for around 15\% of these catalogs.  Thus, if GRB090531B and GRB130716A are indeed extended emission bursts, there are likely an additional $\sim 15$ undetected extended emission bursts classified as long in \swift~with no \fermi~data, and a similar number of undetected extended emission bursts classified as short in \fermi~with no \swift~data.

\subsection{Bulk Properties of Short and Long GRB}

Another way to evaluate whether the separations in these three catalogs are identical is to compare the bulk properties of the short and long populations in all three datasets.  If GRB truly are correctly separated into classes by astrophysical origin, then the short and long GRB observed by each telescope should be drawn from identical distributions, and thus have identical distributions of properties.  Further, with a clean separation between short and long GRB, it should now be possible to determine which other properties correlate with GRB type, something that would not have been possible for complete samples without this method.

Such a comparison is significantly complicated by different selection functions for each telescope.  In particular, the duration is dilated by a factor of $(1+z)$, and therefore telescopes sensitive to GRB at a wider range of redshift should also have a broader distribution of durations.  Because the redshift is only known for a small fraction of GRBs, it is not possible to compare rest-frame durations for the full samples.  

\begin{figure}
    \centering
    \includegraphics[width=0.5\textwidth]{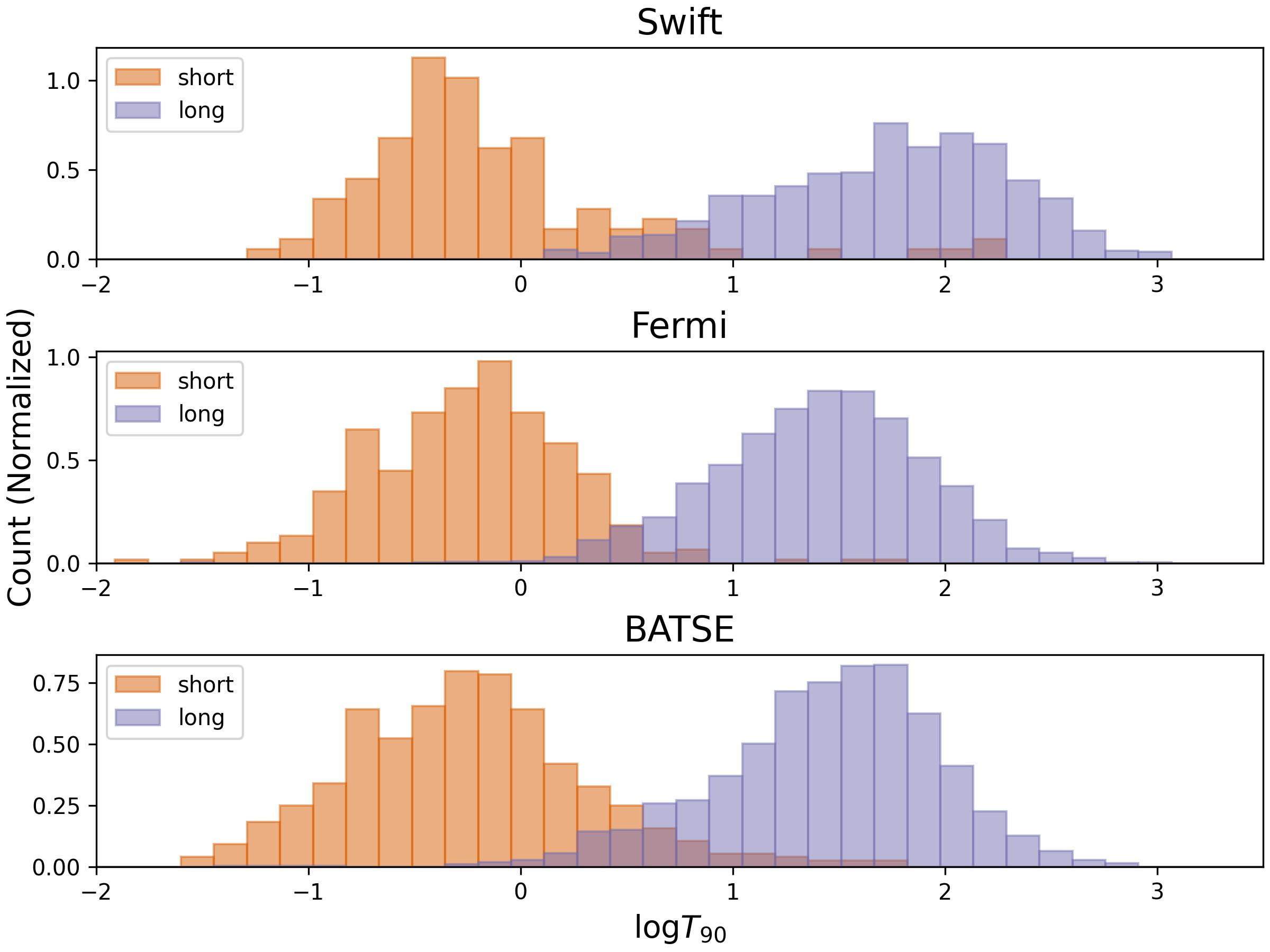}
    \caption{Comparison of the observed duration distributions of short and long bursts in the classifications from the \swift~(top) \fermi~(middle), and\batse~(bottom) datasets.  The redshift distributions in these datasets are expected to be different, although there is insufficient redshift information to verify this.  The distributions of short bursts are qualitatively similar and could even be consistent with having been drawn from very similar distributions if it were possible to correct for time dilation.  However, the \swift~selection of long GRB likely differs from \batse~and \fermi~by more than redshift alone could account for, and is likely due to softer bands providing a different selection than the other two datasets.}
    \label{fig:durationcomp}
\end{figure}

Still, a comparison of the observed durations indicates a similar separation in each dataset (Fig. \ref{fig:durationcomp}).  The most similar, given their similar frequency ranges, are \batse~and ~\fermi.  The short GRB durations are similar in all three datasets, but the long GRBs in \swift include a longer-duration tail than in \batse~or \fermi.  Thus, it is possible that all three telescopes are selecting a similar set of short bursts down to redshift distributions and detection thresholds.  However, the softer bands in \swift produce a dissimilar duration distribution of long bursts when compared with \batse~and \fermi.  

\begin{figure}  
    \centering
    \includegraphics[width=0.45\textwidth]{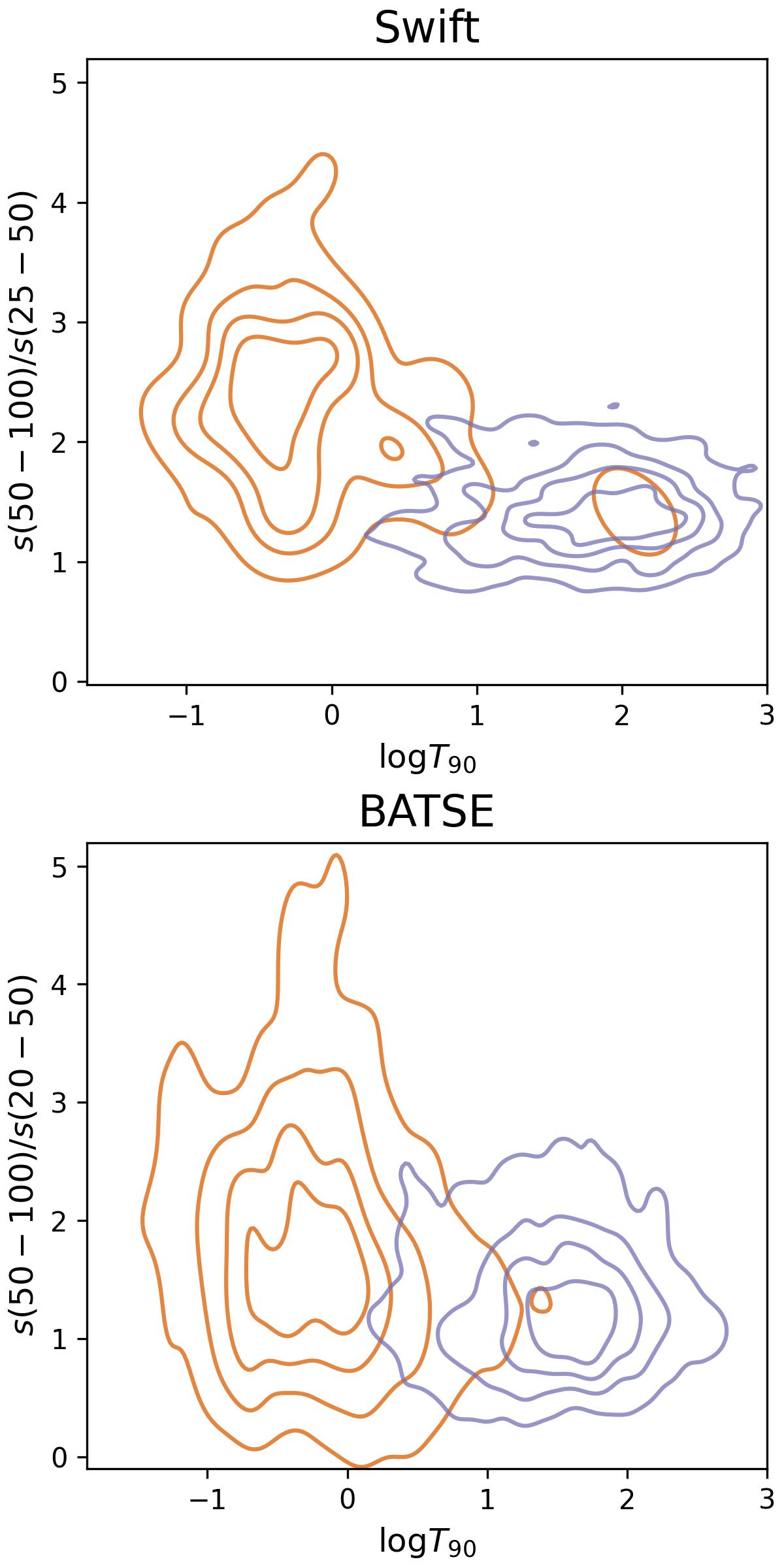}
    \caption{Distribution of objects in hardness and duration in the \swift~(left) and \batse~(right) datasets.  It should be noted that due to different bands used in the calculation, the \swift~and \batse~hardness measurements are on different scales and cannot be compared directly.  The \fermi~catalog does not include a direct measure of hardness.  The two catalogs exhibit a qualitatively similar, overlapping distribution of short and long GRBs.  This is consistent with a selection that depends upon intrinsic GRB properties rather than details of the observatory or processing pipeline.}
    \label{fig:hardnesscomp}
\end{figure}

The distribution of short and long GRBs on a hardness-duration plot further indicates that the t-SNE and UMAP classifications are separating objects with similar bulk properties (Fig. \ref{fig:hardnesscomp}).  Only \batse~and \swift~are shown here, as \fermi~does not produce hardness as part of its data release.  It should be noted that due to different available bands, the hardness measured by \batse cannot be directly compared with that measured by \swift.  For both \batse~and \swift, short GRBs are generally harder and shorter than long GRBs, but with some overlap.  This is again consistent with the hypothesis that short and long GRBs are robust categories intrinsic to GRB emission rather than to details of the observatories or processing pipelines used to measure their properties.

\section{Discussion}
\label{sec:discuss}

Following a pilot study using t-SNE to classify \swift~gamma-ray bursts from the full observed light curves \citep{Jespersen2020}, here dimensionality reduction is shown to be able to classify observations from all three major GRB observatories, \batse~, \swift~, and \fermi.  As with \swift, all three datasets produce a separation into two substantial groups.  This fits with previous work proposing two distinct classes of progenitors: that short bursts are likely associated with mergers \citep{Tanvir2013,berger2013,Ghirlanda2018} and long bursts with the core collapse of massive stars \citep{hjorth2003,Stanek2003}.

In previous studies, the definition of a short or long burst has generally been based on duration only.  Since the two distributions overlap in duration, some bursts will therefore be misclassified.  In this work, the separation into short and long is based on the entire observed light curve, and produces a clean separation into two groups for each dataset without overlap.  The hope is that this clean separation will correspond to physical properties, and that the resulting short and long bursts will indeed have distinct astrophysical origins.  \citet{Jespersen2020} found that a comparison of their classification with both known and proposed supernova afterglows supports this interpretation.

A significant issue with many machine learning methods is that classifications can be based on extraneous information, confounding variables, or metadata.  Thus, the clean separation in the \swift~dataset reported by \citet{Jespersen2020} might occur due to different astrophysical origins, as hoped, but could also be produced by differences in data processing.  A comparison of all three catalogs shows a similar separation with similar properties.  Further, nearly all of the objects observed by both \swift~and \fermi~are classified in the same way.  Therefore, it can be concluded that the separations reported here are truly astrophysical in origin.

\subsection{Additional Classes}

Although 97\% of the bursts common to \swift~and \fermi~are classified identically, 9 are not. Several of these bursts are part of the small fraction with unstable classifications (\S~\ref{tab:outliers}).  However, at least two bursts, GRB090531B and GRB130716A, have entirely stable but conflicting classifications.  In the \swift~data alone, they appear to be typical long bursts, but in the \fermi~data alone, they appear to be typical short bursts.  A strong possibility is that these are part of the previously-reported group of extended emission bursts \citep{Norris2006,Kaneko2015}.  If they have a distinct astrophysical origin, then GRBs should properly be divided into at least three groups, not two.

However, their presence should also suggest that additional classes of bursts might exist which are difficult to classify from any single one of these three observatories.  It might have been hoped that the use of additional information could separate them from more typical short or long bursts.  However, dimensionality reduction algorithms which use all of the available information have been unable to do so.  The authors of this work attempted to tune preprocessing and hyperparameters in order to separate these bursts from the others, but were unable to do so from any single catalog.

Of course, given the combination of the \swift~and \fermi~observations, it is easy to identify these objects as the only ones which are classified differently.  Perhaps this is not so surprising.  Multi-wavelength astronomy has proven to be far more powerful than any single observatory, and multi-messenger astronomy is poised to be similarly powerful.  Thus, a key conclusion here is that the next generation of gamma-ray observatories should be constructed with multi-wavelength observations in mind, and that the combination of \swift~and \fermi has already proven to be more powerful than an improved version of either observatory would be alone.  

It is important to note that ``unsupervised'' algorithms such as the ones used here are in practice strongly dependent on the choice of hyperparameters.  The proper interpretation of the embeddings presented here should not be that the hyperparameters we have chosen are correct and others incorrect.  Rather, every choice of hyperparameters leads to a valid embedding, which contains potentially useful information about the distribution but is always incomplete.  In that sense, it would be more like considering various projections.  Unlike projections, however, there is no rigorous formalism such as principal component analysis for determining which will be most useful.  Thus, an inability in this work to find hyperparameters which separate these possible extended emission bursts from a single catalog does not guarantee that insufficient information exists to do so.

There is also considerable literature on the possibility of a third class consisting of intermediate-duration bursts found in the \batse catalog \citep{Mukherjee1998,Hakkila2003,Horvath2004,Chattopadhyay2007,Ripa2009,Zhang2022}.  Thus, a search for hyperparameters that separate such an intermediate group is well motivated.  Using the same techniques presented here, it was indeed possible to produce an embedding that separates a group of intermediate duration using the \batse 4B catalog.  However, this group does not appear in a catalog restricted to post-4B \batse bursts, and no similar group was found when tuning hyperparameters to search for one in the \fermi~and \swift~catalogs.  As a result, a reasonable conclusion is that this group is not of astronomical origin (cf. \citealt{Hakkila2000}), but rather is related to instrumentation or data reduction techniques applied only to earlier part of the \batse dataset.

\subsection{Physical Interpretations}
\cite{norris2010} suggest that approximately 1/4 of a sample of \swift short GRBs have signatures of extended emission, and estimate that the true fraction may be as high as 50\%. This is not observed in our classifications, which could be due to detector effects, but possibly also due to the "choking" mechanism suggested by \citet{ bucciantini2012}, which would reduce the rate of observed extended emission bursts. All the bursts classified here as short but with $T_{90}>2s$ are within or close to the approximate theoretical boundary of $\approx 100 s$ suggested by \citet{metzger2008, bucciantini2012}.

Although the classification presented here does not immediately allow for distinctions between different progenitor scenarios for the extended emission, there are several signatures that will be exciting to follow with the next generation of GRB observatories. Currently, the only viable way to distinguish between the two primary proposed progenitor scenarios, neutron star - neutron star (NS-NS) mergers and accretion induced collapse (AIC), is by the signature of the elements produced during the event \citep{metzger2008}. This was done for a NS-NS merger by \citet{Watson2019}, who identified strontium in the spectra of the afterglow.  However, these spectra have low signal-to-noise ratios, and need to be taken within a few days, making them hard both to obtain and analyze. 

Another possible way of distinguishing between different short GRB extended emission progenitor mechanisms would be to rely on the extra polarization that would be produced during the prompt emission in the AIC scenario. The proposed \textit{Daksha} mission will carry X-ray polarimeters which will be able to measure the polarization of the prompt emission shortly after a trigger \citep{daksha2022}. Including the polarization would then allow t-SNE/UMAP to further subdivide the short group, corresponding to either NS-NS mergers (without extended emission) or AICs (with extended emission), if both progenitors classes do exist.

This approach would also lend itself to distinguishing between different progenitor mechanisms for LGRBS \citep{toma2009_polarization}, but will require having a large statistical sample due to the large uncertainties in observed prompt emission polarizations \citep{Kole2020POLAR}. For the planned \textit{Daksha} mission, the lowest energy that will be detectable is currently 1 keV, but based on the models of \citet{metzger2008, bucciantini2012}, this should ideally be even lower, preferably past the \swift 0.3 keV limit, in order to best distinguish between different progenitor classes.

\subsection{Robustness of Machine Learning for Astronomical Problems}

It is surprising that even though the embeddings produce clear separations, the classifications are not entirely stable.  Even though the results of \citet{Jespersen2020} appeared unambiguous, and have been reproduced by independent groups from the same codebase, a few objects may still have been misclassified.  Resampling shows that in a study measuring 1000 similar bursts, a small fraction of bursts could end up being classified as either short or long.  Even for the \fermi~catalog, with the most robust classification, 2.0\% of the bursts change classification in at least 10\% of trials.  However, in any individual embedding, there is a clear separation providing an unambiguous assignment of each GRB as either short or long.  

This is one example of a more generic issue when using machine learning methods in astronomy.  Statistical methods typically are associated with theorems proven from a set of axioms.  Thus, it is known with certainty that if those axioms apply to a dataset, a particular method will produce, e.g., the minimum variance unbiased estimator.  Such a method needs a proven theorem in order to be considered valid.

However, machine learning algorithms are often instead validated via benchmark problems and datasets.  An algorithm which outperforms previous attempts at those problems is considered successful.  However, there is rarely a rigorous formalism proving optimality.  One of the advantages of UMAP over t-SNE is that there is a stronger mathematical argument for its embedding.

Further, often these benchmark problems are on idealized, noiseless datasets.  One of the standard dimensionality reduction problems is to classify handwritten digits in the MNIST dataset \citep{mnistlecun1998gradient}.  These digits are noiseless and have no missing pixels.  Indeed, a change as simple as inverting the colors, which leaves the digits equally legible for humans, defeats many state-of-the-art machine learning-based classification schemes \citep{Sun2021}.

However, in scientific applications, both the central value and uncertainty are essential.  Without the latter one cannot determine whether data are consistent with a model.  Dimensionality reduction algorithms do not produce uncertainties in the embedded locations, and there is no widely-adopted standard for doing so.  The resampling method used here was developed by the authors in an attempt to estimate robustness in a non-parametric way.  \footnote{A better approach might have been to construct a large number of simulated datasets based on estimated uncertainties and repeat the procedure for each.  However, even for datasets of this size, the runtime of t-SNE and UMAP is far too long to allow such a Monte Carlo.}

Thus, a significant challenge in using machine learning for astronomical purposes will be developing methods which can properly account for uncertainties.  Rather than turning perfect data into central values, astronomers must turn noisy data with known/estimated uncertainties into central values with known/estimated uncertainties.  As shown in this study, the results can be surprising.  

The authors would like to thank Kasper Heintz and Darach Watson for helpful discussions.  The Cosmic Dawn Center (DAWN) is funded by the Danish National Research Foundation under grant No. 140.  

\bibliographystyle{aasjournal}
\bibliography{ref}

\begin{thebibliography}{}
\expandafter\ifx\csname natexlab\endcsname\relax\def\natexlab#1{#1}\fi
\providecommand{\url}[1]{\href{#1}{#1}}
\providecommand{\dodoi}[1]{doi:~\href{http://doi.org/#1}{\nolinkurl{#1}}}
\providecommand{\doeprint}[1]{\href{http://ascl.net/#1}{\nolinkurl{http://ascl.net/#1}}}
\providecommand{\doarXiv}[1]{\href{https://arxiv.org/abs/#1}{\nolinkurl{https://arxiv.org/abs/#1}}}

\bibitem[{Becht {et~al.}(2019)Becht, McInnes, Healy, Dutertre, Kwok, Ng,
  Ginhoux, \& Newell}]{Becht2019}
Becht, E., McInnes, L., Healy, J., {et~al.} 2019, Nature biotechnology, 37, 38

\bibitem[{{Berger} {et~al.}(2013){Berger}, {Fong}, \& {Chornock}}]{berger2013}
{Berger}, E., {Fong}, W., \& {Chornock}, R. 2013, \apjl, 774, L23,
  \dodoi{10.1088/2041-8205/774/2/L23}

\bibitem[{{Bhalerao} {et~al.}(2022){Bhalerao}, {Vadawale}, \&
  {Tendulkar}}]{daksha2022}
{Bhalerao}, V., {Vadawale}, S., \& {Tendulkar}, S. 2022, in American
  Astronomical Society Meeting Abstracts, Vol.~54, American Astronomical
  Society Meeting Abstracts, 348.13

\bibitem[{{Bromberg} {et~al.}(2011){Bromberg}, {Nakar}, \&
  {Piran}}]{Bromberg11}
{Bromberg}, O., {Nakar}, E., \& {Piran}, T. 2011, \apjl, 739, L55,
  \dodoi{10.1088/2041-8205/739/2/L55}

\bibitem[{Bromberg {et~al.}(2013)Bromberg, Nakar, Piran, \&
  Sari}]{Bromberg_2013classification}
Bromberg, O., Nakar, E., Piran, T., \& Sari, R. 2013, The Astrophysical
  Journal, 764, 179, \dodoi{10.1088/0004-637x/764/2/179}

\bibitem[{{Bucciantini} {et~al.}(2012){Bucciantini}, {Metzger}, {Thompson}, \&
  {Quataert}}]{bucciantini2012}
{Bucciantini}, N., {Metzger}, B.~D., {Thompson}, T.~A., \& {Quataert}, E. 2012,
  \mnras, 419, 1537, \dodoi{10.1111/j.1365-2966.2011.19810.x}

\bibitem[{{Cano} {et~al.}(2017){Cano}, {Wang}, {Dai}, \& {Wu}}]{Cano2017}
{Cano}, Z., {Wang}, S.-Q., {Dai}, Z.-G., \& {Wu}, X.-F. 2017, Advances in
  Astronomy, 2017, 8929054, \dodoi{10.1155/2017/8929054}

\bibitem[{{Chattopadhyay} {et~al.}(2007){Chattopadhyay}, {Misra},
  {Chattopadhyay}, \& {Naskar}}]{Chattopadhyay2007}
{Chattopadhyay}, T., {Misra}, R., {Chattopadhyay}, A.~K., \& {Naskar}, M. 2007,
  \apj, 667, 1017, \dodoi{10.1086/520317}

\bibitem[{Fiedler(1973)}]{Fiedler1973}
Fiedler, M. 1973, Czechoslovak mathematical journal, 23, 298

\bibitem[{{Fruchter} {et~al.}(2006){Fruchter}, {Levan}, {Strolger},
  {Vreeswijk}, {Thorsett}, {Bersier}, {Burud}, {Castro Cer{\'o}n},
  {Castro-Tirado}, {Conselice}, {Dahlen}, {Ferguson}, {Fynbo}, {Garnavich},
  {Gibbons}, {Gorosabel}, {Gull}, {Hjorth}, {Holland }, {Kouveliotou}, {Levay},
  {Livio}, {Metzger}, {Nugent}, {Petro}, {Pian}, {Rhoads}, {Riess}, {Sahu},
  {Smette}, {Tanvir}, {Wijers}, \& {Woosley}}]{06environmentfruchter}
{Fruchter}, A.~S., {Levan}, A.~J., {Strolger}, L., {et~al.} 2006, \nat, 441,
  463, \dodoi{10.1038/nature04787}

\bibitem[{{Ghirlanda} {et~al.}(2018){Ghirlanda}, {Nappo}, {Ghisellini},
  {Melandri}, {Marcarini}, {Nava}, {Salafia}, {Campana}, \&
  {Salvaterra}}]{Ghirlanda2018}
{Ghirlanda}, G., {Nappo}, F., {Ghisellini}, G., {et~al.} 2018, \aap, 609, A112,
  \dodoi{10.1051/0004-6361/201731598}

\bibitem[{{Hakkila} {et~al.}(2003){Hakkila}, {Giblin}, {Roiger}, {Haglin},
  {Paciesas}, \& {Meegan}}]{Hakkila2003}
{Hakkila}, J., {Giblin}, T.~W., {Roiger}, R.~J., {et~al.} 2003, \apj, 582, 320,
  \dodoi{10.1086/344568}

\bibitem[{{Hakkila} {et~al.}(2000){Hakkila}, {Haglin}, {Pendleton}, {Mallozzi},
  {Meegan}, \& {Roiger}}]{Hakkila2000}
{Hakkila}, J., {Haglin}, D.~J., {Pendleton}, G.~N., {et~al.} 2000, \apj, 538,
  165, \dodoi{10.1086/309107}

\bibitem[{{Hjorth} \& {Bloom}(2012)}]{Hjorth2012}
{Hjorth}, J., \& {Bloom}, J.~S. 2012, {The Gamma-Ray Burst - Supernova
  Connection} (Cambridge University Press), 169--190

\bibitem[{{Hjorth} {et~al.}(2003){Hjorth}, {Sollerman}, {M{\o}ller}, {Fynbo},
  {Woosley}, {Kouveliotou}, {Tanvir}, {Greiner}, {Andersen}, {Castro-Tirado},
  {Castro Cer{\'o}n}, {Fruchter}, {Gorosabel}, {Jakobsson}, {Kaper}, {Klose},
  {Masetti}, {Pedersen}, {Pedersen}, {Pian}, {Palazzi}, {Rhoads}, {Rol}, {van
  den Heuvel}, {Vreeswijk}, {Watson}, \& {Wijers}}]{hjorth2003}
{Hjorth}, J., {Sollerman}, J., {M{\o}ller}, P., {et~al.} 2003, \nat, 423, 847,
  \dodoi{10.1038/nature01750}

\bibitem[{{Horv{\'a}th} {et~al.}(2004){Horv{\'a}th}, {M{\'e}sz{\'a}ros},
  {Bal{\'a}zs}, \& {Bagoly}}]{Horvath2004}
{Horv{\'a}th}, I., {M{\'e}sz{\'a}ros}, A., {Bal{\'a}zs}, L.~G., \& {Bagoly}, Z.
  2004, in Astronomical Society of the Pacific Conference Series, Vol. 312,
  Gamma-Ray Bursts in the Afterglow Era, ed. M.~{Feroci}, F.~{Frontera},
  N.~{Masetti}, \& L.~{Piro}, 82

\bibitem[{Hu {et~al.}(2019)Hu, Wang, Hu, Mao, Chen, Yan, Yong, Dong, Wei, Wang,
  {et~al.}}]{Hu2019}
Hu, Y., Wang, X., Hu, B., {et~al.} 2019, PLoS biology, 17, e3000365

\bibitem[{{Jespersen} {et~al.}(2020){Jespersen}, {Severin}, {Steinhardt},
  {Vinther}, {Fynbo}, {Selsing}, \& {Watson}}]{Jespersen2020}
{Jespersen}, C.~K., {Severin}, J.~B., {Steinhardt}, C.~L., {et~al.} 2020,
  \apjl, 896, L20, \dodoi{10.3847/2041-8213/ab964d}

\bibitem[{{Kaneko} {et~al.}(2015){Kaneko}, {Bostanc{\i}},
  {G{\"o}{\u{g}}{\"u}{\c{s}}}, \& {Lin}}]{Kaneko2015}
{Kaneko}, Y., {Bostanc{\i}}, Z.~F., {G{\"o}{\u{g}}{\"u}{\c{s}}}, E., \& {Lin},
  L. 2015, \mnras, 452, 824, \dodoi{10.1093/mnras/stv1286}

\bibitem[{{Kobak} \& {Berens}(2019)}]{Kobak2019}
{Kobak}, D., \& {Berens}, P. 2019, Nature Communications, 10, 5416,
  \dodoi{10.1038/s41467-019-13056-x}

\bibitem[{{Kole} {et~al.}(2020){Kole}, {De Angelis}, {Berlato}, {Burgess},
  {Gauvin}, {Greiner}, {Hajdas}, {Li}, {Li}, {Pollo}, {Produit}, {Rybka},
  {Song}, {Sun}, {Szabelski}, {Tymieniecka}, {Wang}, {Wu}, {Wu}, {Xiong},
  {Zhang}, \& {Zhang}}]{Kole2020POLAR}
{Kole}, M., {De Angelis}, N., {Berlato}, F., {et~al.} 2020, \aap, 644, A124,
  \dodoi{10.1051/0004-6361/202037915}

\bibitem[{{Kouveliotou} {et~al.}(1993){Kouveliotou}, {Meegan}, {Fishman},
  {Bhat}, {Briggs}, {Koshut}, {Paciesas}, \& {Pendleton}}]{kouveliotou1993}
{Kouveliotou}, C., {Meegan}, C.~A., {Fishman}, G.~J., {et~al.} 1993, \apjl,
  413, L101, \dodoi{10.1086/186969}

\bibitem[{LeCun {et~al.}(1998)LeCun, Bottou, Bengio, Haffner,
  {et~al.}}]{mnistlecun1998gradient}
LeCun, Y., Bottou, L., Bengio, Y., Haffner, P., {et~al.} 1998, Proceedings of
  the IEEE, 86, 2278

\bibitem[{{Le{\'s}niewska} {et~al.}(2022){Le{\'s}niewska}, {Micha{\l}owski},
  {Kamphuis}, {Dziadura}, {Baes}, {Cer{\'o}n}, {Gentile}, {Hjorth}, {Hunt},
  {Jespersen}, {Koprowski}, {Floc'h}, {Miraghaei}, {Guelbenzu}, {Oszkiewicz},
  {Palazzi}, {Poli{\'n}ska}, {Rasmussen}, {Schady}, \& {Watson}}]{Lesniewska22}
{Le{\'s}niewska}, A., {Micha{\l}owski}, M.~J., {Kamphuis}, P., {et~al.} 2022,
  \apjs, 259, 67, \dodoi{10.3847/1538-4365/ac5022}

\bibitem[{{Lien} {et~al.}(2016){Lien}, {Sakamoto}, {Barthelmy}, {Baumgartner},
  {Cannizzo}, {Chen}, {Collins}, {Cummings}, {Gehrels}, {Krimm}, {Markwardt},
  {Palmer}, {Stamatikos}, {Troja}, \& {Ukwatta}}]{Lien2016}
{Lien}, A., {Sakamoto}, T., {Barthelmy}, S.~D., {et~al.} 2016, \apj, 829, 7,
  \dodoi{10.3847/0004-637X/829/1/7}

\bibitem[{{McInnes} {et~al.}(2018){McInnes}, {Healy}, \&
  {Melville}}]{McInnes2018}
{McInnes}, L., {Healy}, J., \& {Melville}, J. 2018, arXiv e-prints,
  arXiv:1802.03426.
\newblock \doarXiv{1802.03426}

\bibitem[{{Meegan}(1997)}]{BATSEcatalog}
{Meegan}, C.~A. 1997, {The BATSE Catalog of Gamma-Ray Bursts}, Tech. rep., NASA
  STI

\bibitem[{{Metzger} {et~al.}(2008){Metzger}, {Quataert}, \&
  {Thompson}}]{metzger2008}
{Metzger}, B.~D., {Quataert}, E., \& {Thompson}, T.~A. 2008, \mnras, 385, 1455,
  \dodoi{10.1111/j.1365-2966.2008.12923.x}

\bibitem[{{Mukherjee} {et~al.}(1998){Mukherjee}, {Feigelson}, {Jogesh Babu},
  {Murtagh}, {Fraley}, \& {Raftery}}]{Mukherjee1998}
{Mukherjee}, S., {Feigelson}, E.~D., {Jogesh Babu}, G., {et~al.} 1998, \apj,
  508, 314, \dodoi{10.1086/306386}

\bibitem[{{Nakar}(2007)}]{NAKARsgrb}
{Nakar}, E. 2007, \physrep, 442, 166, \dodoi{10.1016/j.physrep.2007.02.005}

\bibitem[{Ng {et~al.}(2001)Ng, Jordan, \& Weiss}]{Ng2001}
Ng, A., Jordan, M., \& Weiss, Y. 2001, in Advances in Neural Information
  Processing Systems, ed. T.~Dietterich, S.~Becker, \& Z.~Ghahramani, Vol.~14
  (MIT Press).
\newblock
  \url{https://proceedings.neurips.cc/paper/2001/file/801272ee79cfde7fa5960571fee36b9b-Paper.pdf}

\bibitem[{{Norris} \& {Bonnell}(2006)}]{Norris2006}
{Norris}, J.~P., \& {Bonnell}, J.~T. 2006, \apj, 643, 266,
  \dodoi{10.1086/502796}

\bibitem[{{Norris} {et~al.}(2010){Norris}, {Gehrels}, \&
  {Scargle}}]{norris2010}
{Norris}, J.~P., {Gehrels}, N., \& {Scargle}, J.~D. 2010, \apj, 717, 411,
  \dodoi{10.1088/0004-637X/717/1/411}

\bibitem[{{Norris} {et~al.}(1986){Norris}, {Share}, {Messina}, {Dennis},
  {Desai}, {Cline}, {Matz}, \& {Chupp}}]{Norris1986}
{Norris}, J.~P., {Share}, G.~H., {Messina}, D.~C., {et~al.} 1986, \apj, 301,
  213, \dodoi{10.1086/163889}

\bibitem[{{Paciesas} {et~al.}(1999){Paciesas}, {Meegan}, {Pendleton}, {Briggs},
  {Kouveliotou}, {Koshut}, {Lestrade}, {McCollough}, {Brainerd}, {Hakkila},
  {Henze}, {Preece}, {Connaughton}, {Kippen}, {Mallozzi}, {Fishman},
  {Richardson}, \& {Sahi}}]{paciesas1999}
{Paciesas}, W.~S., {Meegan}, C.~A., {Pendleton}, G.~N., {et~al.} 1999, \apjs,
  122, 465, \dodoi{10.1086/313224}

\bibitem[{{Stanek} {et~al.}(2003){Stanek}, {Matheson}, {Garnavich}, {Martini},
  {Berlind}, {Caldwell}, {Challis}, {Brown}, {Schild}, {Krisciunas}, {Calkins},
  {Lee}, {Hathi}, {Jansen}, {Windhorst}, {Echevarria}, {Eisenstein}, {Pindor},
  {Olszewski}, {Harding}, {Holland }, \& {Bersier}}]{Stanek2003}
{Stanek}, K.~Z., {Matheson}, T., {Garnavich}, P.~M., {et~al.} 2003, \apjl, 591,
  L17, \dodoi{10.1086/376976}

\bibitem[{{Sun} {et~al.}(2021){Sun}, {Zeng}, \& {Zhang}}]{Sun2021}
{Sun}, Y., {Zeng}, Y., \& {Zhang}, T. 2021, iScience, 24, 102880,
  \dodoi{10.1016/j.isci.2021.102880}

\bibitem[{{Tanvir} {et~al.}(2013){Tanvir}, {Levan}, {Fruchter}, {Hjorth},
  {Hounsell}, {Wiersema}, \& {Tunnicliffe}}]{Tanvir2013}
{Tanvir}, N.~R., {Levan}, A.~J., {Fruchter}, A.~S., {et~al.} 2013, \nat, 500,
  547, \dodoi{10.1038/nature12505}

\bibitem[{Tavani {et~al.}(1998)Tavani, Kniffen, Mattox, Paredes, \&
  Foster}]{tavani1998}
Tavani, M., Kniffen, D., Mattox, J., Paredes, J., \& Foster, R. 1998, The
  Astrophysical Journal Letters, 497, L89

\bibitem[{{Toma} {et~al.}(2009){Toma}, {Sakamoto}, {Zhang}, {Hill},
  {McConnell}, {Bloser}, {Yamazaki}, {Ioka}, \&
  {Nakamura}}]{toma2009_polarization}
{Toma}, K., {Sakamoto}, T., {Zhang}, B., {et~al.} 2009, \apj, 698, 1042,
  \dodoi{10.1088/0004-637X/698/2/1042}

\bibitem[{{von Kienlin} {et~al.}(2020){von Kienlin}, {Meegan}, {Paciesas},
  {Bhat}, {Bissaldi}, {Briggs}, {Burns}, {Cleveland}, {Gibby}, {Giles},
  {Goldstein}, {Hamburg}, {Hui}, {Kocevski}, {Mailyan}, {Malacaria},
  {Poolakkil}, {Preece}, {Roberts}, {Veres}, \&
  {Wilson-Hodge}}]{VonKienlin2020}
{von Kienlin}, A., {Meegan}, C.~A., {Paciesas}, W.~S., {et~al.} 2020, \apj,
  893, 46, \dodoi{10.3847/1538-4357/ab7a18}

\bibitem[{{{\v{R}}{\'\i}pa} {et~al.}(2009){{\v{R}}{\'\i}pa},
  {M{\'e}sz{\'a}ros}, {Wigger}, {Huja}, {Hudec}, \& {Hajdas}}]{Ripa2009}
{{\v{R}}{\'\i}pa}, J., {M{\'e}sz{\'a}ros}, A., {Wigger}, C., {et~al.} 2009,
  \aap, 498, 399, \dodoi{10.1051/0004-6361/200810913}

\bibitem[{{Watson} {et~al.}(2019){Watson}, {Hansen}, {Selsing}, {Koch},
  {Malesani}, {Andersen}, {Fynbo}, {Arcones}, {Bauswein}, {Covino}, {Grado},
  {Heintz}, {Hunt}, {Kouveliotou}, {Leloudas}, {Levan}, {Mazzali}, \&
  {Pian}}]{Watson2019}
{Watson}, D., {Hansen}, C.~J., {Selsing}, J., {et~al.} 2019, \nat, 574, 497,
  \dodoi{10.1038/s41586-019-1676-3}

\bibitem[{Xiang {et~al.}(2021)Xiang, Wang, Yang, Wang, Xu, \& Chen}]{Xiang2021}
Xiang, R., Wang, W., Yang, L., {et~al.} 2021, Frontiers in genetics, 12, 646936

\bibitem[{Zhang {et~al.}(2012)Zhang, Shao, Yan, \& Wei}]{zhang2012revisiting}
Zhang, F.-W., Shao, L., Yan, J.-Z., \& Wei, D.-M. 2012, The Astrophysical
  Journal, 750, 88

\bibitem[{{Zhang} {et~al.}(2022){Zhang}, {Shao}, {Zhang}, {Zou}, {Sun}, {Yao},
  \& {Li}}]{Zhang2022}
{Zhang}, S., {Shao}, L., {Zhang}, B.-B., {et~al.} 2022, \apj, 926, 170,
  \dodoi{10.3847/1538-4357/ac4753}

\end{thebibliography}

\appendix

A data table containing the classification of each burst is given here and available online in a machine-readable format.  For each burst, a classification is given for each telescope which observed. The possible classifications are the following types:

\begin{itemize}
    \item S (short): a burst identified as short by both t-SNE and UMAP and in at least 90\% of resampled t-SNE maps.
    \item L (long): a burst identified as long by both t-SNE and UMAP in at least 90\% of resampled t-SNE maps.
    \item A (ambiguous): a burst which was classified as both short and long in at least 10\% of resampled t-SNE maps.
    \item D (disagreement): a burst which is classified differently by t-SNE and UMAP, but which is classified consistently in at least 90\% of resampled t-SNE maps.
\end{itemize}

In addition, there is an overall classification given for each burst.  If the burst is observed by only one telescope, the overall classification is same as for that telescope.  For bursts observed by both \swift~and \fermi, if both give the same classification, this also the overall classification.  A burst which is S or L in one catalog and A or D in the other is classified as S or L based on the unambiguous classification.  The three bursts which are classified as L by \swift but S by \fermi, GRB090531B, GRB090927, and GRB130716A, are classified as D (disagreement) in the overall classification.  These may be extended emission bursts, as discussed in the main text.

The primary name used for each burst the name given each individual catalog, with the \fermi~name used for objects observed by both \fermi~and \swift.  Probabilities given are the probability that any given burst is short in resampled data.

\begin{table*}[!ht]
\begin{center}
 \begin{tabular}{llccccccc} 
 \hline
Name & \fermi~ Name & Type & \batse Type & \batse Prob. & \swift Type & \swift Prob. & \fermi Type & \fermi Prob.\\
 \hline
GRB910421 &  & L & L & 0.0 &  &  &  & \\
GRB910423 &  & L & L & 0.0 &  &  &  & \\ 
GRB910424 &  & L & L & 0.01 &  &  &  & \\
GRB910425A &  & L & L & 0.0 &  &  &  & \\ 
GRB910425B &  & L & L & 0.0 &  &  &  & \\ 
GRB910426 &  & L & L & 0.0 &  &  &  & \\ 
GRB910427 &  & L & L & 0.0 &  &  &  & \\ 
GRB910429 &  & L & L & 0.0 &  &  &  & \\ 
GRB910430 &  & L & L & 0.0 &  &  &  & \\ 
GRB910501 &  & L & L & 0.0 &  &  &  & \\ 
$\cdots$ \\
GRB090927 & GRB090927422 & D &  &  & L & 0.07 & S & 1.0 \\
GRB090928 & GRB090928646 & L &  &  &  &  & L & 0.01 \\
GRB090929 & GRB090929190 & L &  &  &  &  & L & 0.0 \\
GRB090929A &  & L &  &  & L & 0.01 &  & \\ 
GRB090929B &  & L &  &  & L & 0.0 &  & \\ 
GRB091002 & GRB091002685 & L &  &  &  &  & L & 0.02 \\
GRB091003 & GRB091003191 & L &  &  &  &  & L & 0.0 \\
GRB091005 & GRB091005679 & L &  &  &  &  & L & 0.0 \\
GRB091006 & GRB091006360 & S &  &  &  &  & S & 0.99 \\
$\cdots$ \\
 \hline
\end{tabular}
\caption{Classifications based on this work for GRBs in the \batse, \swift, and \fermi~catalogs.  Bursts with missing data or metadata which were excluded from the analysis are not included in the table.  A full, machine-readable version is available online.}
\label{tab:catalog}
\end{center}
\end{table*}

\end{document}